\documentclass[twocolumn,aps,longbibliography,superscriptaddress,showpacs]{revtex4-1}
\usepackage{setspace}

\newcommand{\btab}{\begin{tabular}}
\newcommand{\etab}{\end{tabular}}

\newcommand{\rank}{{\bf rank}}

\newcommand{\trace}{{\bf Tr}}

\newcommand{\diag}{{\mrm diag}}

\newcommand{\eps}{\varepsilon}

\newcommand{\alf}{\alpha}
\newcommand{\om}{\omega}

\renewcommand{\th}{\theta}
\newcommand{\del}{\delta}
\newcommand{\sig}{\sigma}

\newcommand{\Del}{\Delta}
\newcommand{\Gam}{\Gamma}
\newcommand{\Om}{\Omega}

\newcommand{\norm}[1]{ \left\| #1 \right\| }

\newcommand{\Ecal}{{\mathcal E}}

\newcommand{\eg}{\emph{e.g.}}
\newcommand{\ie}{\emph{i.e.}}

\newcommand{\bquem}{\begin{quote}\begin{em}}
\newcommand{\equem}{\end{em}\end{quote}}

\newcommand{\blist}{\begin{description}}
\newcommand{\elist}{\end{description}}

\newcommand{\bquote}{\begin{quote}}
\newcommand{\equote}{\end{quote}}

\newcommand{\ben}{\begin{enumerate}}
\newcommand{\een}{\end{enumerate}}

\newcommand{\bit}{\begin{itemize}}
\newcommand{\eit}{\end{itemize}}

\newcommand{\bea}{\begin{array}}
\newcommand{\eea}{\end{array}}

\newcommand{\bds}{\begin{displaystyle}}
\newcommand{\eds}{\end{displaystyle}}

\newcommand{\Rbf}{{\mathbf R}}
\newcommand{\Cbf}{{\mathbf C}}

\newcommand{\ds}{\displaystyle}

\newcommand{\refeq}[1]{(\ref{eq:#1})}

\newcommand{\refsec}[1]{\ref{sec:#1}}

\makeatletter
\def\beq{\@ifnextchar 
[{\@tempswatrue\@beq}{\@tempswafalse\@beq[]}}
\def\@beq[#1]{\begin{equation}\edef\@tmparg{#1}\ifx\@tmparg\@e
mpty \else
	\label{#1}\fi}
\newcommand{\eeq}{\end{equation}}
\makeatother 

\newcommand{\beqaa}{\begin{eqnarray*}}
\newcommand{\eeqaa}{\end{eqnarray*}}

\newcommand{\beqa}{\begin{eqnarray}}
\newcommand{\eeqa}{\end{eqnarray}}

\newcommand{\bc}{\begin{center}}
\newcommand{\ec}{\end{center}}

\newcommand{\bfig}{\begin{figure}}
\newcommand{\efig}{\end{figure}}

\renewcommand{\diag}{\mbox{\rm diag}}

\newcommand{\schrod}{Schr\"{o}dinger\ }

\renewcommand{\rank}{\mbox{\rm rank}}

\usepackage[colorlinks=true,
            urlcolor=blue,
            citecolor=red,
            pdfstartview=FitH,
            pdfpagemode=None]{hyperref}

\newcommand{\black}[1]{\textcolor{black}{#1}}

\usepackage{times}
\usepackage{graphicx}
\usepackage{amssymb}
\usepackage{amsmath}
\usepackage{amsfonts}
\usepackage[dvips]{epsfig}

\usepackage{bbm}

\makeatletter
\newcommand*\rel@kern[1]{\kern#1\dimexpr\macc@kerna}
\newcommand*\widebar[1]{%
  \begingroup
  \def\mathaccent##1##2{%
    \rel@kern{0.8}%
    \overline{\rel@kern{-0.8}\macc@nucleus\rel@kern{0.2}}%
    \rel@kern{-0.2}%
  }%
  \macc@depth\@ne
  \let\math@bgroup\@empty \let\math@egroup\macc@set@skewchar
  \mathsurround\z@ \frozen@everymath{\mathgroup\macc@group\relax}%
  \macc@set@skewchar\relax
  \let\mathaccentV\macc@nested@a
  \macc@nested@a\relax111{#1}%
  \endgroup
}
\makeatother

\newcommand{\figref}[1]{Figure~\ref{fig:#1}}
\newcommand{\cmag}{c_{\rm mag}}
\newcommand{\comp}{compression}
\newcommand{\comps}{compressions}

\newcommand{\mtb}{{m_{\rm tbw}}}
\renewcommand{\aa}{\alf}
\newcommand{\lin}{{\rm lin}}
\newcommand{\Slin}{S_\lin}
\newcommand{\Rlin}{R_\lin}
\newcommand{\Shlin}{{\widebar{S}_\lin}}
\newcommand{\Rhlin}{\widebar{R}_\lin}
\newcommand{\Lam}{\Lambda}
\newcommand{\delf}{\Del}
\newcommand{\Htfi}{H_{\rm tfi}}
\newcommand{\Hnldi}{{{\cal H}_{\rm NLDI}}}

\newcommand{\sinc}{{\rm sinc}}
\newcommand{\sv}{{\sigma}}

\newcommand{\nspin}{{n_{\rm spin}}}

\newcommand{\U}{\mbox{\bf U}}
\newcommand{\mat}[1]{
\begin{bmatrix}
#1
\end{bmatrix}
}

\newcommand{\Gpinv}{{G^\#}}
\newcommand{\tht}{{\th_t}}

\newcommand{\normfro}[1]{\norm{#1}_{\rm fro}}

\renewcommand{\Rbf}{\mathbb{R}}
\renewcommand{\Cbf}{\mathbb{C}}

\renewcommand{\trace}{\mbox{\bf Tr}}
\renewcommand{\rank}{\mbox{rank }}

\newtheorem{thm}{Theorem}

\renewcommand{\th}{\theta}

\begin{document}

\title{
Quantum System Compression: A Hamiltonian Guided Walk Through Hilbert Space. 
}


\author{Robert L. Kosut}
\affiliation{SC Solutions, Sunnyvale CA, 94085}
\affiliation{Department of Chemistry,
  Princeton University, Princeton, NJ, 08544}


\author{Tak-San Ho}
\author{Herschel Rabitz}
\affiliation{Department of Chemistry,
  Princeton University, Princeton, NJ, 08544}

\date{\today}

\begin{abstract}  
\black{We present a systematic study of quantum system compression for
  the evolution of generic many-body problems. The necessary numerical
  simulations of such systems are seriously hindered by the
  exponential growth of the Hilbert space dimension with the number of
  particles.} For a \emph{constant} Hamiltonian system of Hilbert
space dimension $n$ whose frequencies range from $f_{\min}$ to
$f_{\max}$, we show \black{via a proper orthogonal decomposition},
that for a run-time $T$, the dominant dynamics are compressed in the
neighborhood of a subspace whose dimension is the smallest integer
larger than the time-bandwidth product
$\delf=(f_{\max}-f_{\min})T$. We also show how the distribution of
initial states can further compress the system dimension.  Under the
stated conditions, the time-bandwidth estimate reveals the
\emph{existence} of an effective compressed model whose dimension is
derived solely from system properties and not dependent on the
particular implementation of a {\color{black}variational} simulator,
such as a machine learning system, or quantum device. However, finding
an efficient solution procedure \emph{is} dependent on the simulator
implementation{\color{black}, which is not discussed in this paper}.
In addition, we show that the compression rendered by the proper
orthogonal decomposition encoding method can be further strengthened
via a \black{multi-layer} autoencoder. Finally, we present numerical
illustrations to affirm the compression behavior in time-varying
Hamiltonian dynamics in the presence of external fields. {\color{black}We also
  discuss the potential implications of the findings for machine
  learning tools to efficiently solve the many-body or other high
  dimensional Schr{\"o}dinger equations.}

\end{abstract}  

\maketitle

\section{Introduction}

\black{Numerous recent studies
  \cite{CarleoTroyer:17,Czischek:2018,Schmitt:2018,Fabiani:2019,
    Sarma:2019,Carleo:2019RMP}
utilizing a flexible representation of a variational quantum state
have been proposed based on artificial neural networks (ANN) for
solving many-body quantum problems.  These studies have shown a favorable
polynomial scaling with respect to the system's number of
particles. Such findings
are consistent across many
fields~\cite{Lloyd:1996,illusion:11,sloppy:2013,Freericks:2014,cheap:17,Deep:2020};
despite system complexity in the underlying physics, much of observed
behavior is compressed, \ie, the dominant dynamics is manifested in
significantly lower dimensions.  Compression arises also in the search
over the quantum control landscape as a favorable scaling of control
complexity~\cite{RussellRW:17,KosutArenzRabitz:2019}.}

By compression of quantum system dynamics we mean a simulator that has
these two features: (1) the simulator has a reduced number of
variables that do not scale exponentially with the number of quantum
particles (or, the number of simulator variables is exponentially
smaller than the Hilbert space dimension), and (2) the state error,
upon using the simulator, remains satisfactory for the intended
purpose. In this context, we present a \emph{time-bandwidth product}
which reveals the existence of a compressed system which satisfies the
stated features.  Naturally the level of compression in (1) above is
inversely related to the degree of dynamical error tolerated in (2).
The title of the paper can be understood, since the compression occurs
in the system's Hilbert space guided by particular characteristics of
the Hamiltonian involved as shown in the main body of the paper.  The
key time-bandwidth product is reminiscent of ``Hartley's Law''
\cite{Hartley:1928}-\footnote{Hartley's Law: ``It is shown that when
  the storage of energy is used to restrict the steady state
  transmission to a limited range of frequencies the amount of
  information that can be transmitted is proportional to the product
  of the width of the frequency-range by the time it is available.''
} referenced in \cite{LloydM:2014} in relation to optimal control
complexity. This potential compression of quantum dynamics depends
only on the system properties, {\color{black} including} the range of
{\color{black}the eigenvalues of Hamiltonian}, the simulation run-time,
and the initial state.
Though the compression is not dependent on the particular method of
simulation, achieving a similar level of compression is expected to be
dependent on the simulator implementation.  {\color{black}Additionally
  we show that the compression rendered by proper orthogonal
  decomposition can be further reduced via a \black{multi}-layer
  autoencoder. We remark that the compressibility analysis in this
  paper will mainly be carried out for constant Hamiltonians
%
  as well as the autoencoder technique for estimating the reduced
  dimensionality. Numerical illustrations for time- varying Hamiltonian dynamics
  in external (control) fields will also be presented to support the
  pervasive nature of system compression behavior.  Evidence further
  supporting this result comes from the tested Hamiltonians ranging
  from many-body coupled spin systems to those chosen randomly.  }

The paper is written in a style of introducing concepts along with the
associated mathematical formulation as well as clarifying numerical
illustrations throughout the text to best express the various aspects
of quantum system compression as they naturally arise.  The paper is
organized as follows: Section ~\ref{sec:var} introduces the
variational state problem, {\color{black}while}
Section~\ref{sec:linvar} describes the method of proper orthogonal
decomposition {\color{black} for obtaining linear variational states}.
Section~\ref{sec:lti} focuses on constant Hamiltonian dynamics with an
example provided in Section~\ref{sec:exlti}. Sections~\ref{sec:prop}
and \ref{sec:linswp} present the main theoretical results followed by
particular numerical simulation tests in Section~\ref{sec:numerical}.
The utilization of an autoencoder to enhance compression is presented
in Section~\ref{sec:ae}. Additional numerical illustrations for
time-varying dynamics in external fields are presented in
Section~\ref{sec:tv dyn}, and we present extensions on compression for
unitary dynamics and for nonlinear frequency sweeps in
Section~\ref{sec:extend}. A discussion of the findings in the paper is
given in Section~\ref{sec:sum}. Finally, details of particular
derivations are given in the Appendix.

\section{Variational state}
\label{sec:var}
The goal is to simulate the quantum state $\psi_t\in\Cbf^n$ for
$t\in[0,T]$ by a \emph{variational quantum state} $x(\tht)\in\Cbf^n$,
a function of a time-varying parameter $\tht\in\Cbf^m$ where $m<n$. We
refer to $\tht$ as the \emph{compressed state} and to $m$ as
\emph{compression}, respectively. The {\color{black}effective} ``small
volume in Hilbert space'' referred to in \cite{illusion:11} and
exposed in \cite{sloppy:2013} is akin to the {\color{black}compressed}
state discussed here.  In a machine learning system, such as a neural
network, the ``parameters'' are the weights that connect all the
layers. These weighting paramters are not necessarily the same as the
{\color{black}compressed dimension or the associated} state variables,
though it is possible depending on the simulator implementation. In
this note we leave the simulator parametrization and implementation
unspecified and seek to show that generally there are fewer effective
compressed states ($m$) than that of the full Hilbert space ($n$).
Irrespective of the implementation, it is assumed that the variational
state $x(\tht)$ is a smooth function of the time-varying parameter
{\color{black}vector} $\tht$, and as a result,
\beq[eq:grad]
\dot x(\tht)= G(\tht)\dot\tht,
\quad
G(\th) = \nabla_\th x(\th) \in\Cbf^{n\times m},
\eeq
where $G(\tht)$ as indicated is a gradient matrix.  

\subsection{Variational formulations}
\label{sec:VP}

In the most general case the system to be simulated is the
time-varying quantum system,
\beq[eq:psit]
i{\dot\psi_t} = H_t{\psi_t},\ t\in[0,T].
\eeq
The data available for simulation is the initial state
$\psi_0\in\Cbf^n$ and the time-varying Hamiltonian
$H_t\in\Cbf^{n\times n}$, a description that encompasses a quantum
system under a known control or external field. The variational
{\color{black} optimization problem that we \emph{seek to solve}} is to
minimize the following \emph{functional},
\beq[eq:Eth]
E[\th] =
\ds
\frac{1}{T}
\int_0^T
\norm{\psi_t-x(\tht)}_2^2dt.
\eeq
{\color{black}The qualifing phrase \emph{``seek to solve''} is stated
  because} the actual state flow $\{\psi_t,t\in[0,T]\}$ is not
available in the simulation context. If it were
{\color{black}available} then there is no need for a variational
version unless one seeks to {\it\color{black} post facto} find a low
dimensional representation.  In general, minimizing $E[\th]$ is not
possible without the actual state. As in \cite{CarleoTroyer:17} for
variational simulation, there is a means to query as necessary the
available data: the initial state $\psi_0$ and the time-varying
Hamiltonian $\{H_t,t\in[0,T]\}$. The variational problem is then
pragmatically posed to minimize a functional such as,
\beq[eq:Eqth]
\Ecal[\th] =
\ds
\frac{1}{T}
\int_0^T
\big\|H_tx(\tht)-iG(\tht)\dot\tht)\big\|_2^2~dt.
\eeq
%
Depending on the context we will
refer to $E[\th]$ or its integrand as {\color{black}the} \emph{state
  error}, and to $\Ecal[\th]$ or its integrand as {\color{black}the}
\emph{equation error}. To construct a variational quantum simulator in
a classical device using only the available data (\eg,
$\psi_0\in\Cbf^n,\{H_t\in\Cbf^{n\times n},t\in[0,T]\}$) is equivalent
to minimizing an equation error functional such as \refeq{Eqth}. This
is clearly a necessary surrogate for the ideal goal: minimizing the
state error \refeq{Eth}.

\subsection{\color{black}Compression}
\label{sec:Perfection}

If the variational solution is perfect then $x(\tht)=\psi_t,\forall
t\in[0,T]$ from which it follows that both state and equation errors
(and associated functionals) are zero. As a consequence the
variational parameter $\tht$ must satisfy,
\beq[eq:thopt]
G(\tht)\dot\tht + iH_t\psi_t =0,
\quad
x(\th_0)=\psi_0,
\quad
t\in[0,T].
\eeq
These equations for finding $\dot\tht\in\Cbf^m$ with $m<n$ are
over-determined, \ie, more equations ($n$) than variables ($m$). As a
result, \refeq{thopt} will hold only if $H_t\psi_t$ (or $i\dot\psi_t$)
is a linear combination of the columns of $G(\th_t)$. In particular, a
solution for $\dot\tht$ to solve \refeq{thopt} will exist if and only
if for $t\in[0,T]$,
\beq[eq:GGpinv]
\bea{l}
\left(I_m-\Gam_m(\tht)\right)H_t\psi_t=0,
\\
\Gam_m(\tht) = G(\tht)\Gpinv(\tht)\in\Cbf^{n\times n}.
\eea
\eeq
where $\Gpinv(\tht)\in\Cbf^{m\times n}$ is the \emph{pseudo-inverse}
of $G(\tht)$. Note that properties of the pseudo-inverse yields
$\Gam_m(\tht)^2=\Gam_m(\tht)$, \ie, it is idempotent.
Condition \refeq{GGpinv} is necessary and sufficient for the
variational state to provide a perfect simulation, \ie, $x(\tht)$ and
true state $\psi_t$ are identical.  


For a perfect simulation, and clearly for an imperfect but very good
simulation, especially with $m\ll n$, or perhaps more importantly,
where $m$ does not scale exponentially with the number of particles as
does $n$, the state flow $\psi_t$ from \refeq{psit} \emph{must be
  inherently compressible}.  To understand {\color{black} the basic
  foundations of quantum system compression for now} we focus on a
\emph{time-invariant} quantum system, meaning that $H_t=H$ is a
constant matrix{\color{black}; the time-dependent extension will be
  returned to in Section~\ref{sec:tv dyn}.}  In addition we restrict
attention to a \emph{linear} variational state, that is, the
variational state gradient is a constant matrix:
$G(\tht)=M\in\Cbf^{n\times m},\ m<n$. {\color{black} The introduction
  of nonlinearity into the compression process will be treated with an
  autoencoder in Section~\ref{sec:ae}.}  Assuming that the eigenvalues
sweep linearly {\color{black}from their lowest to highest values}, we
show that it is {\color{black} the} system properties alone which
allow the state and equation error measures to show a favorable
scaling of \comp\ $m$ with system dimension $n$. Our analysis provides
a means to assess the \comp\ range that might be achievable from a
more generally flexible variational state parametrization, \ie, with a
time-varying nonlinear gradient $G(\tht)$. As in numerous studies
showing favorable compression, we are led to the assertion that a
variational quantum solution of a time-invariant quantum system is
possible with its compression dependent solely on the system
parameters: Hamiltonian eigenvalues, initial state, and simulation run
time. However, the demonstrated \emph{existence} of system compression
shown in this paper still leaves the challenge to explicitly exploit
that for the performance of dynamical simulations (\ie, especially for
high dimensional many-body situations ) in, for example, a machine
learning format.

\section{Linear Variational State}\label{sec:linvar}

A common means of obtaining a \emph{linear} variational state is the
method of \emph{proper orthogonal decomposition} (POD) or equivalently
{\color{black}often referred to as} \emph{principal component analysis}
(PCA) \cite{Kutz:2013}. There are several variants of POD, here we use an
unbiased version where the variational state is set to $x(\tht)=M\tht$
with gradient matrix $M\in\Cbf^{n\times m}$ where $m<n$ \footnote{In
  POD/PCA often a bias term is included: $x(\tht)=M\tht+b$.}. The
optimization variables $M$ and $\tht$ are selected to minimize a state
error functional {\color{black}chosen as} \refeq{Eth}:
\beq[eq:pod opt]
\bea{ll}
\mbox{minimize} &\ds
E_{\rm pod} = \frac{1}{T}\int_0^T\norm{\psi_t-M\tht}_2^2 dt,
\\
\mbox{subject to} & M\in\Cbf^{n\times m},\
\tht\in\Cbf^m,t\in[0,T].
\eea
\eeq
The well known optimal solution is,
\beq[eq:pod soln]
\tht = M_m^\dag\psi_t,
\quad
M_m = \arg\min_{M\in\Cbf^{n\times m}}\ \trace(I_n-MM^\dag)C[\psi],
\eeq
where $M_m\in\Cbf^{n\times m}$ is found via a singular value
decomposition of the $n\times n$ positive semi-definite ``covariance''
matrix,
\beq[eq:cov]
\bea{rcl}
C[\psi] &=& \frac{1}{T}\int_0^T \psi_t\psi_t^\dag dt
\\
&=&
\mat{M_m&M_{n-m}}
\mat{Q_m&0\\0&Q_{n-m}}
\mat{M_m&M_{n-m}}^\dag,
\eea
\eeq
with singular values arranged (as usual) in descending order in
$Q_m=\diag(\sv_1(C),\ldots,\sv_m(C)$ and
$Q_{n-m}=\diag(\sv_{m+1}(C),\ldots,\sv_n(C))$ and where
$\mat{M_m&M_{n-m}}\in{\bf U}(n)$. The resulting optimal (POD)
variational state is,
\beq[eq:xpod]
x_t = \Gam_m\psi_t,\
\Gam_m = M_mM_m^\dag\in\Cbf^{n\times n},
\eeq
which produces the minimum state POD error measure: the sum of the
smallest $n-m$ singular values of the covariance matrix, \ie,
\beq[eq:Epodmin]
E_{\rm pod}=\trace(I_n-\Gam_m)C[\psi])
=\sum_{i=m+1}^n\sv_i(C).
\eeq
Because the matrix $M_m$ is part of a unitary matrix, its $m$ columns are
orthonormal vectors in $\Cbf^n$ and thus $M_m^\dag M_m=I_m$. The
columns of $M_m$ form a basis set for the {\color{black}compressed}
state $\tht$ making $\Gam_m$ idempotent, \ie, $\Gam_m^2=\Gam_m$.
Combining \refeq{xpod} with \refeq{Eqth} we get the corresponding POD
equation error measure,
\beq[eq:Eqpodmin]
\Ecal_{\rm pod} 
=\frac{1}{T}\int_0^T\norm{(H_t\Gam_m-\Gam_mH_t)\psi_t}_2^2dt.
\eeq
A typical measure of model reduction error is the relative error in
the cumulative sum of singular values of the covariance matrix
compared to the sum of all the singular values:
\beq[eq:sumsvC]
\epsilon_m(C) = 1 - \sum_{i=1}^m\sv_i(C)/\sum_{i=1}^n\sv_i(C)
=\sum_{i=m+1}^n\sv_i(C).
\eeq
%
To predict compression quantitatively for a quantum system, we make an
obvious restriction as discussed next which leads to the
time-bandwidth product presented in Section~\ref{sec:linswp}. Later in
Section~\ref{sec:tv dyn} we will discuss the compression effect with a
time-varying Hamiltonian and suggest how a compression estimate can be
obtained.

\section{Time-Invariant Hamiltonian Dynamics}\label{sec:lti}

Consider a \emph{time-invariant} {\color{black} Hamiltonian driving the
  quantum dynamical} system,
\beq[eq:lti]
i\dot\psi_t = H\psi_t,\ t\in[0,T].
\eeq
Since the Hamiltonian $H\in\Cbf^{n\times n}$ is constant, the standard
eigenvalue decomposition yields,
\beq[eq:eigH]
H = V\Om V^\dag
\quad
\left\{
\bea{l}
V\in{\bf U}(n),
\\
\Om=\diag(\om),
\quad
\om\in\Rbf^n.
\eea
\right.
\eeq
Subsequently the state flow $\psi_t$ can be expressed variously as,
\beq[eq:eigpsit]
\bea{rcl}
\psi_t &=& Ve^{-it\Om}V^\dag\psi_0
= Vf_t, 
\\
f_t &=& \mat{\aa_1e^{-i\om_1t}&\cdots&\aa_ne^{-i\om_nt}}^T,
\\
\aa &=& V^\dag\psi_0, 
\eea
\eeq
where $\aa\in\Cbf^n,\ \norm{\aa}_2=1$ is the projection of the initial
state onto the natural basis of the Hamiltonian. As shown in Appendix
\ref{app:svR}, the singular values of the POD covariance matrix
\refeq{cov} are identical with those of a positive semidefinite matrix
$R\in\Cbf^{n\times n}$ which has the decomposition,
\beq[eq:RSdecomp]
\bea{l}
R = \diag(\aa)~S~\diag(\aa)^\dag,
\\
S  = \sinc(\Lam),
\\
\Lam_{k\ell}=(\om_k-\om_\ell)T/2,\ k,\ell=1,\ldots,n.
\eea
\eeq
We will refer to $R$ as a ``covariance'' matrix and to $S$ as the
``sinc'' matrix with respective elements,
\beq[eq:RSkell]
\bea{rcl}
{R}_{k\ell}
&=&
\left\{
\bea{l}
|\aa_k|^2,\ k=\ell,
\\
\ds
(\aa_k\aa_\ell^*)~\sinc~\Lam_{k\ell},\
k\neq\ell,
\eea
\right.
\\
S_{k\ell} &=&
\left\{
\bea{l}
1,\ k=\ell,
\\
\ds \sinc~\Lam_{k\ell},\
k\neq\ell.
\eea
\right.
\eea
\eeq
The corresponding normalized singular value errors in the form of
\refeq{sumsvC} for $m=1,\ldots,n$ are,
\beq[eq:esvRS]
\bea{rcl}
\epsilon_m(R) &=& \ds 1-\sum_{i=1}^m\sv_i(R),
\\
\epsilon_m(S) &=& \ds 1-\sum_{i=1}^m\sv_i(S)/n,
\eea
\eeq
where the cumulative sum of singular value errors are normalized by
their respective singular value sums:
$\trace~R=\norm{\aa}_2^2=\norm{\psi_0}_2^2=1$ and $\trace~S=n$.
Though the calculation of the elements of $R$ and $S$ do \emph{not}
require any dynamical simulation, to use these expressions
necessitates obtaining an eigenvalue decomposition of the system
Hamiltonian, knowledge of the initial state, and calculating the
singular values of $R$ and $S$. We can, however, without doing any
such decompositions, assert some generic properties of compression.
Before discussing these, it is worthwhile to do an example.

\section{Example: Time-invariant Hamiltonian spin system}\label{sec:exlti}

\begin{figure}[t]
  \centering
\renewcommand{\arraystretch}{0.2}
\btab{c}
  \includegraphics[height=1.5in,width=\columnwidth]{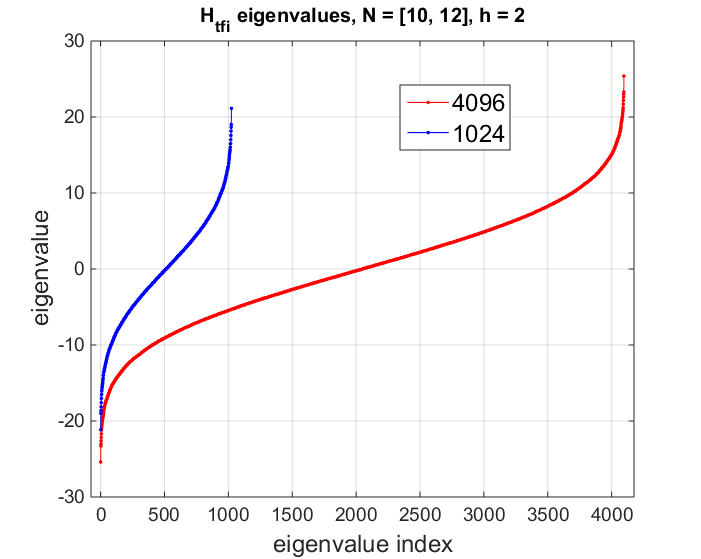}
  \\(a)\\
  \includegraphics[height=2.4in,width=\columnwidth]{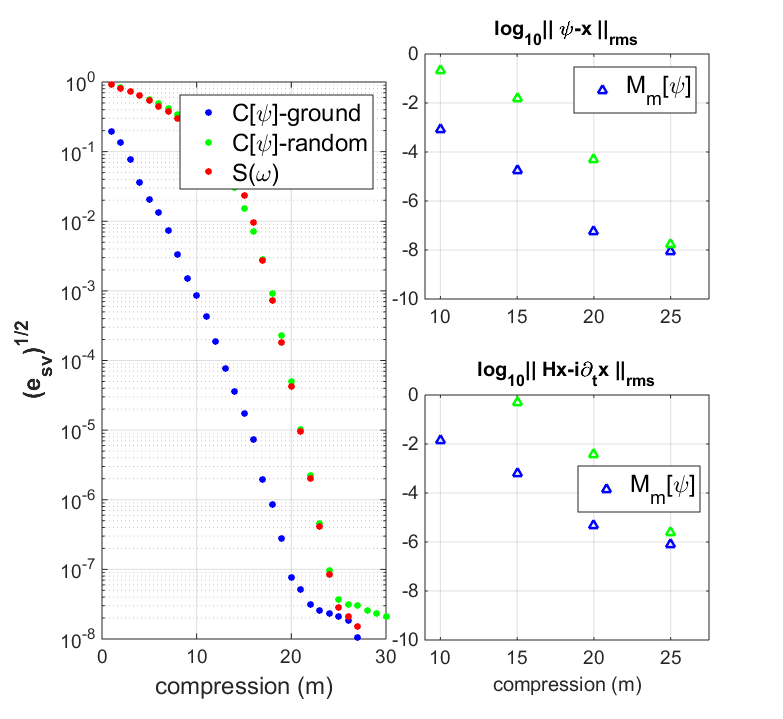}
 \\(b)\\
 \includegraphics[height=2.4in,width=\columnwidth]{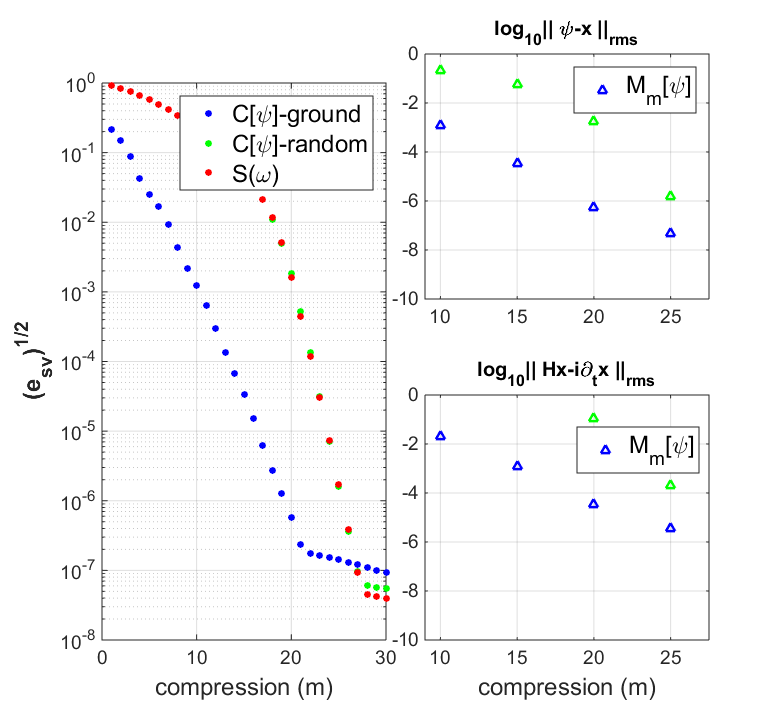}
 \\(c)
 \etab
\caption{{\bf TFI}. All runs with $h=2$ \refeq{Htfi}, run-time $T=2$,
  10 and 12 spins (eigenvalues in {\bf (a)}) and two initial states:
  one at the ground state of $\Htfi$ with $h=4$ (blue dots and
  triangles in {\bf (b)-(c)}; one randomly selected (green dots and
  triangles in {\bf (b)-(c)}). The left sub-plots in {\bf (b)-(c)}
  show the relative RMS error (square-root of \refeq{esvRS}) of the
  covariance matrix $R$ (blue dots) and the sinc matrix $S$ (red dots)
  from \refeq{RSkell} for \comps\ $m=1,\ldots,30$. Lower right
  sub-plots in {\bf (b)-(c)} show errors for $m\in\{10,15,20,25\}$. }
  \label{fig:htfi}
  \end{figure}

\begin{figure}[t]
  \centering
  \includegraphics[width=0.75\columnwidth]{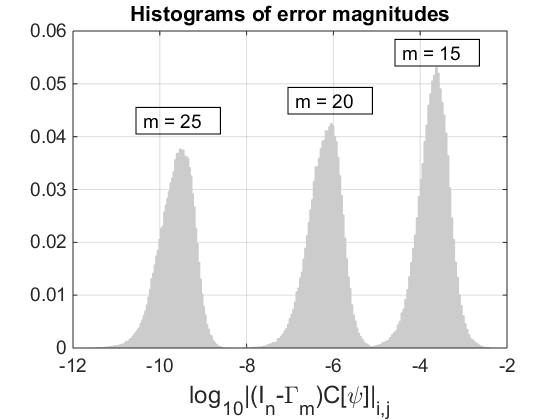}
  \caption{{\bf Histograms of state errors} The log of the absolute
    values of all elements of state error matrix $(I_n-\Gam_m)C[\psi]$
    from \refeq{Epodmin} for $n=2^{10}$ and for the selected \comps\
    shown.}
  \label{fig:hist}
\end{figure}

Consider the transverse-field Ising (TFI) $\nspin$-system with
Hamiltonian,
\beq[eq:Htfi]
H_{\rm TFI} = -h\sum_{i=1}^\nspin\sig^x_i
- \sum_{i=1}^{\nspin-1}\sig^z_i \sig^z_{i+1}.
\eeq
This system was studied in \cite{CarleoTroyer:17}; here we use the
same model parameters and initial state choices but only for a limited
number of spins. The upper plot {\bf (a)} in \figref{htfi} shows the
eigenvalues of $H_{\rm TFI}$ with TFI parameter $h=2$ for
$\nspin\in\{10,12\}$, ergo, space dimension
$n=2^\nspin\in\{1024,4096\}$. The eigenvalues clearly sweep uniformly
from max to min and sum to zero, thus forming two groups of $n/2$
positive eigenvalues and $n/2$ negative eigenvalues with each group
having the same magnitudes:
$\om\in\{-\om_1,\ldots,-\om_{n/2},\om_1,\ldots,\om_{n/2}\}$.

Plots {\bf(b)-(c)} in \figref{htfi} correspond to the two spin
examples: $\nspin\in\{10,12\}$ each with run time $T=2$. In each of
the {\bf (b)-(c)} plots, the two sub-plots on the right show,
respectively, the RMS values over $t\in[0,T]$ of both state error and
equation error, \ie, $\norm{\psi-x}_{\rm rms}$ and $\norm{Hx-i\dot
  x}_{\rm rms}$. (These are the square-roots, respectively, of $E_{\rm
  pod}$ and $\Ecal_{\rm pod}$ from \refeq{Epodmin}-\refeq{Eqpodmin}.)
The left sub-plots of {\bf(b)-(c)} show the the relative error in the
cumulative sum of singular values of the covariance matrix $R$ (blue
dots) and the sinc matrix $S$ (red dots), both calculated from
\refeq{RSkell}. For comparison with the RMS error measures we use the
square root of the cumulative singular value errors of $R$ and $S$
from \refeq{esvRS}, namely, $\sqrt{\epsilon_m(R)}$ and
$\sqrt{\epsilon_m(S)}$.  The error measures are all computed for
$\Htfi$ with $h=2$, and for each of two initial states: one fixed at
the ground state of $\Htfi$ with $h =4$ (blue dots), and one with the
initial state randomly selected (green dots). (Only results for the
ground state initialization was presented in \cite{CarleoTroyer:17}.)
As seen in \figref{htfi} the selection of a random initial state
(green dots in the singular value plots and green triangles in the error
measure plots) increases the variational \comp\ to achieve the same
relative model error, though the increases are not dramatic.

\figref{hist} displays the TFI data in histograms of $\log_{10}$
of the absolute values of all elements of state error matrix
$(I_n-\Gam_m)C[\psi]$ from \refeq{Epodmin} for $n=2^{10}$ and for the
selected \comps\ shown. The decrease in error magnitudes is quite
dramatic for \comps\ greater than 15.  This value is effectively
predicted by the time-bandwidth product, \ie,
$\delf=(f_{\max}-f_{\min})T\approx 14$.

Clearly there is significant dynamic compression: not only is the
level small compared to the system dimension, moreover, there is
little increase in \comp\ with increasing number of spins. In both
cases shown a \comp\ of order $m=25$ suffices to produce very small
state and equation errors.

\begin{figure}[t]
  \centering
  \btab{cc}
  \includegraphics[width=0.5\columnwidth]{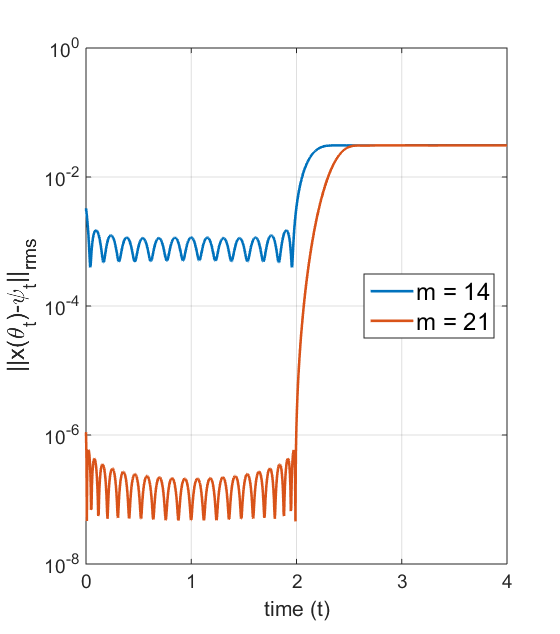}
  &
  \includegraphics[width=0.5\columnwidth]{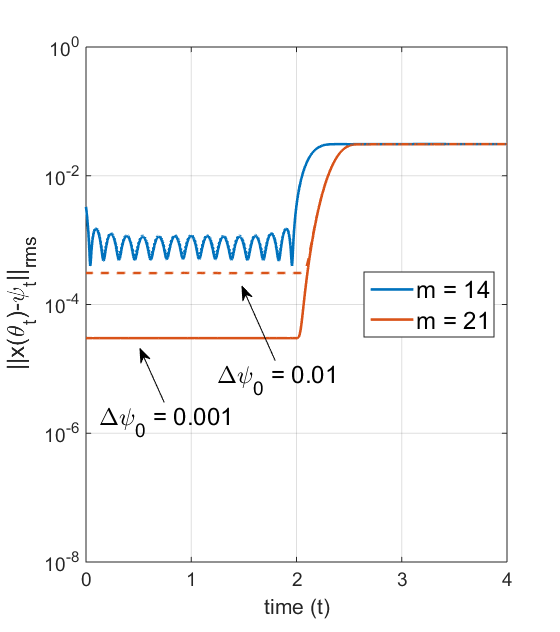}
  \\
  {\bf (a)} & {\bf (b)} 
  \etab
  \caption{{\bf Robustness of RMS magnitude of POD error time
      responses} {\bf(a)} change to run time $T=4$ from variational
    states optimized for run time $T=2$, and {\bf(b)} small
    perturbations to initial state.}
  \label{fig:val}
\end{figure}

The compression order selections $(m)$ are based on the initial state
and system dynamics over the specified simulation time interval
$t\in[0,T]$ with $T=2$. The left plot {\bf(a)} in \figref{val} shows
what happens when the run time is extended to $T=4$ for
\comps\ $m\in\{14,21\}$. As expected, outside of the design range of
$t\in[0,2]$ the state RMS error as a function of time dramatically
increases. In the right plot {\bf(b)} the initial state differs from
that used in the SVD of the covariance matrix, and as a result the
state RMS error deteriorates over the whole time interval. Two levels
of initial state perturbation are shown:
$\Del\psi_0\in\{0.01,0.001\}$. For $m=14$ there is no significant
change whereas for $m=21$ the change is considerable, more than two
orders of magnitude increase in error. The difference in robustness to
initial state may be attributed to the nominal error magnitude, \ie,
the state error for $m=14$ is much larger than that for $m=21$, and
hence, the latter is more sensitive to changes in the initial state.

\section{Run-time and Singular value bounds}
\label{sec:prop}
In the previous examples the relative singular value error of $S$, the
sinc-matrix, provides an upper bound on the errors of the covariance
matrix $R$, and these errors are close when the initial state is
random. This suggests that the efficacy of a variational quantum state
of a time-invariant quantum system \refeq{lti} can be obtained from
the singular values of $R$ (equivalently $C$) and the sinc-matrix $S$
\refeq{RSdecomp}-\refeq{RSkell}. To see this we first examine some
qualitative properties at the extremes of run time $T$.

\subsection{Run-time} 
\label{sec:Runtime}

In the limit as the run-time $T$ goes to infinity the sinc matrix
approaches the identity, hence,
\beq[eq:svR Tinf]
\lim_{T\to\infty}\sv(R)=\sv(\diag(|\aa|^2).
\eeq
The singular values of $R$ become the sorted magnitudes of $\aa$, the
initial state expressed via the Hamiltonian eigenvectors. These are
the diagonal elements of $R$, all the other elements tend to
zero. Compression to $m<n$ in this case requires that the last $n-m$
elements of $\aa$ are much smaller than the first $m$ elements.  At
the opposite end, as the run-time $T$ goes to zero, $R\to
\aa\aa^\dag$, thus,
\beq[eq:svR Tzero]
\lim_{T\to 0}\sv(R)
=
\left\{\sum_{k=1}^n|\aa_k|^2,0,\ldots,0\right\}=\{1,0,\ldots,0\}.
\eeq
Note that there is only one non-zero singular value at
$\norm{\aa}_2=\norm{\psi_0}_2=1$. A reduction to a model with
dimension one is certainly extreme, but expected. The message to take
here is that the dynamics becomes further compressed as run-time
decreases.  In effect, for small run-times not very much of the space
gets filled out beyond where the state started.

The extreme run-time scenarios reflect the fact that since
$\aa=V^\dag\psi_0$, compression depends on how the initial state is
projected onto the eigenvectors of the Hamiltonian. In the previous
examples (\figref{htfi}) when the initial state is prepared
{\color{black} nearby the} ground state, the error measure for $R$ is
significantly smaller than for $S$ because the initial state is
localized in the Hilbert space. For a random initial state we see that
the singular value errors for $R$ and $S$ are close since the initial
state now is more spread out.

\subsection{Singular value bounds} 
\label{sec:SVB}

More quantitative insights can be revealed using a standard singular
value inequality for the product of matrices. From the relation of $R$
and $S$ \refeq{RSdecomp}, the singular values of $R$ are bounded by,
\beq[eq:svRbnd]
\sv_k(R) \leq
\min\Big\{
\norm{\aa}^2_\infty\sv_k(S),\ \norm{S}\sv_k^2(\aa)
\Big\},\quad
k=1,\ldots,n. 
\eeq
%
Here we use the notation $\sv_k(\aa)$ to mean
$\sv_k(\diag~\aa)$, so $\sv_k(\aa)$ assumes that the elements of $\aa$
have been reordered in deceasing magnitude. Another well known
inequality follows, namely,
\beq[eq:rankR]
\rank R \leq \min\Big\{\rank{S},\ \rank{\aa} \Big\}.
\eeq
This rank inequality is only useful if $S$ has some zero singular
values and/or $\aa$ has some zero elements. The latter can occur when
the initial state lies completely in a lower dimensional subspace,
\eg, \cite{KumarMohan:2014}. Though this is generally unlikely, as
observed in the previous spin system examples, compression is possible
because many singular values of $R$ are nearly zero, possibly driven
by the localized initial state lying dominantly being in a low
dimensional subspace. For example, suppose that no elements of $\aa$
are zero and that $\sv_k(S)\approx 0$ for $k>m$ with $m\ll n$. Then
$\sv_k(R)\approx 0$ for $k>m$, and thus $m$ is the maximum variational
\comp. In this case the compression order is bounded by the number of
non-zero singular (or non-small) singular values of $S$, the sinc
matrix.  This property is clearly seen in the numerical results of
\figref{htfi} with the TFI example. We see this in many other cases
that we have run: the sinc-matrix error bounds the covariance matrix
error, \ie, $\epsilon_m(R)\leq\epsilon_m(S)$, or equivalently,
\beq[eq:SleqR]
\sum_{i=1}^m\sv_i(S)/n \leq \sum_{i=1}^m\sv_i(R).
\eeq
Both sides of this inequality are norms, specifically Ky Fan $m$-norms
respectively of $S/n$ and $R$. From the \emph{Ky Fan Dominance
  Theorem} \cite{HornJohnson} the above will hold for all
$m=1,\ldots,n$ if and only if $\norm{S/n}\leq\norm{R}$ for any
{\color{black} norm invariant unitary transformation. Only in} very few
cases have we seen this {\color{black} behavior} violated.
%
%
Nevertheless, at the moment \refeq{SleqR} remains a sufficient
condition for a general frequency sweep. For a \emph{linear} sweep
approximation we can make stronger statements.


\section{Linear frequency sweep}\label{sec:linswp}

\begin{figure}[t]
  \centering
  \btab{cc}
  \includegraphics[width=0.5\columnwidth]{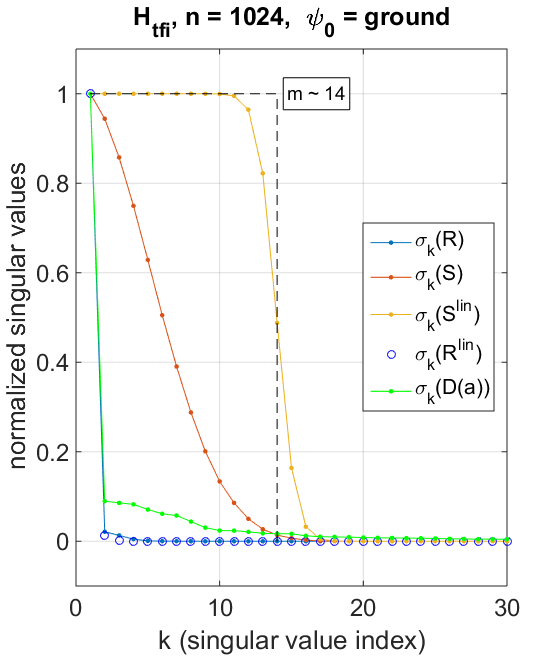}
  &
  \includegraphics[width=0.5\columnwidth]{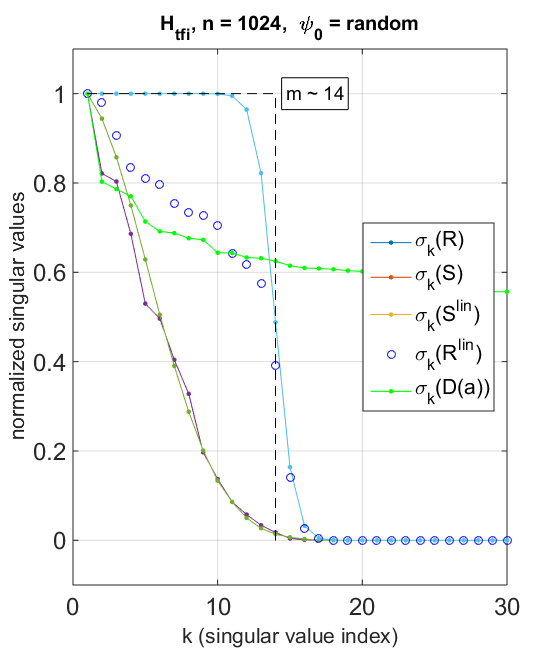}
  \\
    {\bf(a)} & {\bf(b)}
    \\
  \includegraphics[width=0.5\columnwidth]{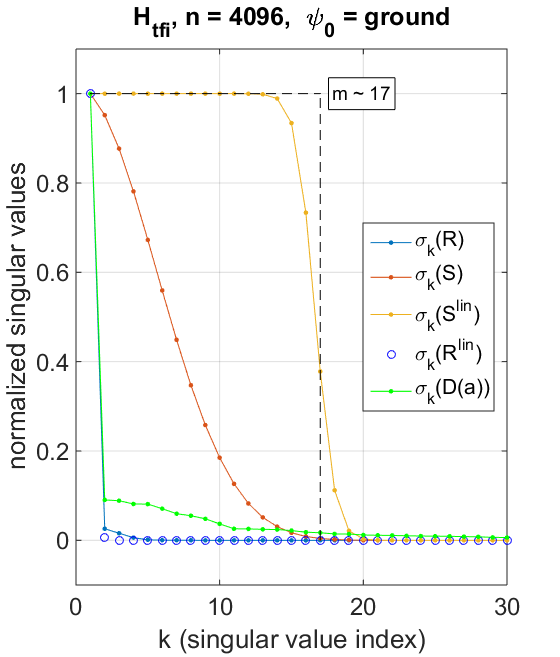}
  &
  \includegraphics[width=0.5\columnwidth]{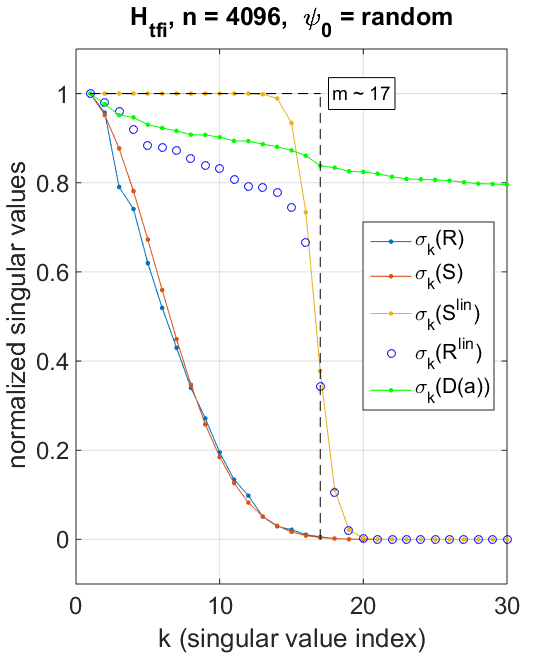}
  \\
  {\bf(c)} & {\bf(d)}
  \etab
  \caption{{\bf Normalized singular values} of $R$, $S$ and $\Slin$
    for $\Htfi$ \refeq{Htfi} with $h=2$ for 10 and 12 spins and with
    initial states at ground for $h=4$ ({\bf(a,c)}) and random
    ({\bf(b,d)}). The rect-function (dashed-lines) is $\bar{S}_\lin$
    from \refeq{svShlin} with associated singular values as indicated
    by the text boxes $\mtb=14$ and $\mtb=17$.}
   \label{fig:svSSlin}
\end{figure}

In many quantum systems, such as just observed for the spin system
example, the eigenvalues of the Hamiltonian are almost linear from
minimum to maximum, or reasonably approximated as such over a large
portion of the range (see \figref{htfi}). Under the assumption
of a linear sweep of eigenvalues, the sinc matrix $S$ \refeq{RSdecomp}
becomes a symmetric (real) \emph{Toeplitz} matrix,
\beq[eq:delf]
\bea{rcl}
\Slin &=& \sinc~\Lam^{\rm lin}\in\Rbf^{n\times n},
\\
\Lam_{kl}^{\rm lin} &=& (\om_k-\om_\ell)T/2 =
\left(\frac{k-\ell}{n-1}\right)\delf\pi,\
k,\ell=1,\ldots,n,
\\
\delf &=& (\om_{\max}-\om_{\min})T/2\pi.
\eea
\eeq
The non-dimensional variable $\delf$, referred to here as the
``time-bandwidth product,'' will be seen to play the key role in
establishing an approximate upper bound on the variational model
order.  The covariance matrix corresponding to $\Slin$ follows from
the form of \refeq{RSdecomp},
\beq[eq:rlin]
\Rlin=\diag(\aa)~\Slin~\diag(\aa)^\dag,
\eeq
{\color{black}in which} $\Rlin$ is Hermitian matrix but not generally
of Toeplitz form. The spin system results in \figref{svSSlin} are of
sufficiently small size so that the singular values of the $R$, $S$,
$\Slin$, and $\Rlin$ matrices can be directly calculated on a standard
laptop.  For large dimensions, this becomes infeasible. Fortunately,
even for very large $n$, the singular values of $\Slin$ can be
approximated by taking the digital Fourier transform (DFT) of a column
of a related \emph{circulant} matrix, for which there are a number of
versions, all resulting in asymptotic approximations of the
eigenvalues \cite{GS:58,Gray:72}. A similar procedure was utilized in
\cite{inelastic:75}. In the case here, with a very specific (sinc)
function forming the symmetric Toeplitz matrix elements, we can appeal
to a more direct result in \cite{Bottcher:2017,Ekstrom:17} on the
asymptotic distribution of the Toeplitz eigenvalues.  As shown in
Appendix \ref{app:svS}, for large $n$ the singular values of the
Toeplitz sinc matrix $\sv_i(\Slin) ,i=1,\ldots,n$, are well
approximated by,
\beq[eq:svSlin]
\sv_i(\Slin) \approx
\left\{
\bea{ll}
(n-1)/\delf, & i < \mtb,
\\
0, &  i \geq \mtb,
\eea
\right.
\eeq
where the index value $\mtb$, under the conditions of a linear
frequency sweep, \emph{is} the level of compression,
\beq[eq:m]
\mtb = \Big\lceil\frac{n}{n-1}\delf\Big\rceil,\quad
\delf =(f_{\max}-f_{\min})T.
\eeq
The frequency spread is expressed here in Hz using $f=2\pi\om$.  The
corresponding covariance error \refeq{esvRS} is,
\beq[eq:esvSlin]
\bea{rcl}
\epsilon_k(\Slin)
&=&
1-\sum_{i=1}^k\sv_i(\Slin)/\sum_{i=1}^n\sv_i(\Slin)
\\
&\approx&
\left\{
\bea{ll}
1-k/m & k < \mtb,
\\
0 &  k \geq \mtb.
\eea
\right.
\eea
\eeq
%
\black{Note that one could multiply the constant Hamiltonian $H$ by $a$
  and divide the run-time $T$ by $a$ and get the same value for
  $\mtb$. This is a trivial scaling for a constant Hamiltonian,
  however, as we will see later (Section~\refsec{tv dyn}), this simple
  scaling does not apply for a time-varying Hamiltonian where $\mtb$
  is interpreted differently to account for the observed compression.}

At the \comp\ level $\mtb$, the variational state errors are
either very small or rapidly decreasing for $m>\mtb$. For quantum
systems with $n\gg 1$ we can take $\mtb=\lceil\delf\rceil$. Let
$\Shlin$ denote the ideal sinc matrix with exactly
$\mtb=\lceil\delf\rceil$ non-zero constant singular values,
\beq[eq:svShlin]
\sig_k(\Shlin)=\left\{
\bea{ll}
n/\delf, & k < \lceil\delf\rceil,
\\
0, & k \geq \lceil\delf\rceil.
\eea
\right.
\eeq
Let $\Rhlin$ denote the corresponding covariance matrix,
\beq[eq:Rhlin]
\Rhlin = \diag(\aa)~\Shlin~\diag(\aa)^\dag.
\eeq
Application of the singular value bound \refeq{svRbnd}
results in,
\beq[eq:Rhlinbnd]
\bea{rcll}
\sig_k(\Rhlin) &\leq&
(n/\delf)\sig_k^2(a), & k < \lceil\delf\rceil,
\\
&=& 0, & k \geq \lceil\delf\rceil.
\eea
\eeq
This {\color{black} result} ensures that for a linear frequency sweep
and large $n$ the Ky Fan Dominance Theorem holds for the ideal pair
$(\Rhlin,\Shlin)$, namely, $\sig_k(\Shlin/n)\leq\sig_k(\Rhlin)$
because $\norm{\aa}_\infty^2\geq 1/n$ since $\norm{\aa}_2=1$.  Thus
for a linear sweep, the ideal relative covariance error is bounded by
the relative sinc error.  Further, if for all elements
$|\aa_i|^2=1/n,\ i=1,\ldots,n$, then both the singular values of
$\Rhlin$ and $\Shlin$ as well as their relative errors coincide.  This
tendency was observed for random initial states in the spin example
and is seen more pointedly in the numerical simulations presented
next.

{\color{black} Note that} the time-bandwidth product $\mtb$ is not
independent of the system dimension $n$. For example, the spin system
Hamiltonian \refeq{Htfi} frequency range is \emph{linear} in the
number of spins $\nspin$, and since $n=2^\nspin$, it follows that
$\mtb\sim\log n$. More specifically,
\beq[eq:mlogn]
\mtb = \Big\lceil(\log n/\log n_0)\delf_0\Big\rceil
\eeq
where $\Del_0,n_0$ are chosen nominal values for comparison with $n\geq
n_0$. For example, with $n_0=2^{10}$ and $m_0=\lceil\delf_0\rceil$, if
$n=2^{20}$ then $\mtb=2m_0$, or if $n=2^{80}$ then $\mtb=8m_0$ and so
on. More generally, if the frequency range scales as a polynomial
function of the number of two-level particles, say $\sim N^k$, then
$\mtb\sim(\log n)^k$.
{\color{black}Such behavior is still a very} favorable scaling with
respect to the exponential scaling of the system dimension with the
number of quantum particles.

\section{Numerical results}
\label{sec:numerical}

\subsection{TFI system}
\label{sec:TFI}

\figref{svSSlin} shows plots of the normalized singular values of $R$,
$S$, $\Slin$ and $\Shlin$ for $\Htfi$ \refeq{Htfi} with $h=2$ for 10
and 12 spins and with two initial states: {\bf(a,c)} at the ground
state for $h=4$ and {\bf(b,d)} random.  Both $R$ and $S$ are
calculated using the actual (nonlinear) eigenvalue sweep of $\Htfi$
whereas $\Slin$ uses a linear sweep over the same range. The text
boxes with $\mtb=14$ and $\mtb=17$ show the ideal predicted \comp\ for
the two spin cases assuming a linear frequency sweep. The ratio
$17/14\approx 1.21$ follows the log-scaling \refeq{mlogn} with $\log
4096/\log 1024 = 1.2$ which obviously is the ratio of spins $12/10$.
For 80 spins the \comp\ estimate increases by a factor of 8 over 10
spins to $m=112$, dramatically low compared to $n=2^{80}$, and so on. The
dashed line rect-function next to these boxes is the ideal singular
value function given by \refeq{eigSlin}: a constant until index $k=m$
and then drops to zero thereafter. The \comp\ level where the singular
values of \emph{both} $S$ and $\Slin$ drop to essentially zero is
almost identical to that predicted by \refeq{m}. Additionally, the
\emph{actual} singular values of $\Slin$ follow those of the ideal
$\Shlin$, tending more closely as $n$ increases from 1024 to 4096.

The break points (\ie, compression level) where the singular values drop
significantly as predicted by the sinc matrix $S$, or the linear sweep
matrix $\Slin$, can differ, and exceed, those of the covariance matrix
$R$ (or $\Rlin$), whose break points are generally smaller especially
from a ground state. They adhere closely to $S$ (or $\Slin$) for a
random initial state. The former is to be expected considering that
the sinc matrix, $S$, contains no information about any correlations
with the initial state.  We also see that the linear frequency sweep
approximation, resulting in the symmetric Toeplitz matrix $\Slin$ and
the corresponding covariance $\Rlin$, are in agreement with these
findings.

\begin{figure}[t]
  \centering
  \includegraphics[width=0.75\columnwidth]{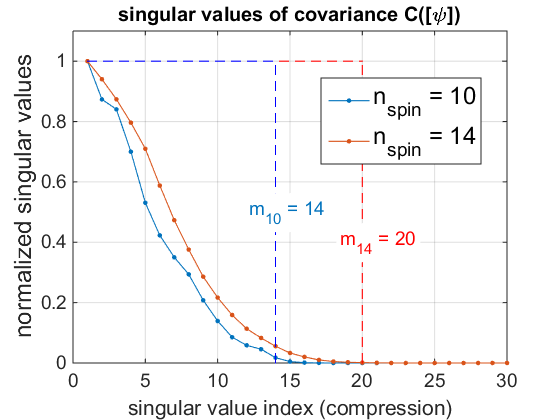}
  \caption{{\bf Singular values} of snapshot matrix \refeq{snap} for
    $\Htfi$ system with $\nspin\in\{10,14\}$ each initialized from a
    random state.}
  \label{fig:svpsi14}
\end{figure}

\subsection{POD via Snapshot}
\label{sec:POD}

\figref{svpsi14} shows a further comparison with $\nspin\in\{10,14\}$
where the singular values are computed from the ``snapshot'' version
of the covariance matrix \refeq{cov},
\beq[eq:snap]
\Psi = \mat{\psi_{t_1}&\cdots&\psi_{t_{K}}}\in\Cbf^{n\times K}
\eeq
where $K=200$ uniformly spaced time samples over the same simulation
run-time $T=2$ as in the previous TFI examples. This gives a sampling
rate many times the maximum Hamiltonian frequency. For both spin
settings the dynamics are initialized with a random state.  The
singular values shown are the squares of the singular values of the
$n\times 200$ snapshot matrix $\Psi$; these approximate those of the
covariance matrix, \ie, $C([\psi])\approx(T/K)\Psi\Psi^\dag$. The
time-bandwidth predicted \comp\ levels $\mtb\in\{14,20\}$ are again
validated by the data and also follow the log scaling \refeq{mlogn}
for these spin systems, \ie, \comp\ ratio $20/14\approx 1.43$ compared
to spin ratio $14/10=1.4$.

\begin{figure}[t]
  \centering
  \includegraphics[width=\columnwidth]{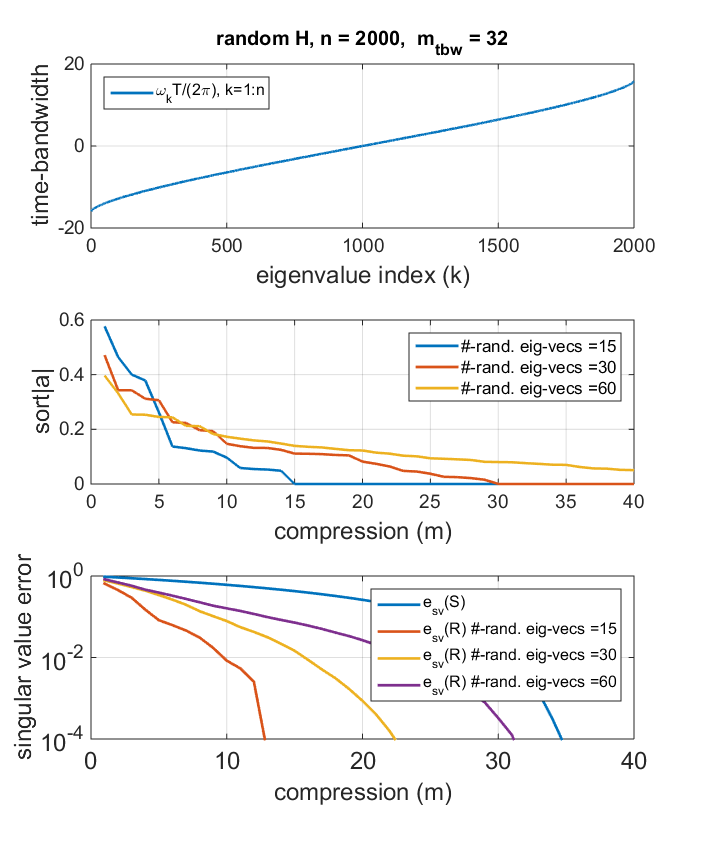}
  \caption{{\bf Random Hamiltonian} n=2000, initial state confined to
    random selection of 15, 30, and 60 eigenvectors of Hamiltonian.}
  \label{fig:randH}
\end{figure}

\subsection{Random Hamiltonians}
\label{sec:Random}

We generated many random Hamiltonians for $n=2000$ and with a variety
of distributions of elements. The results always come down to
confirming the largest compression value in the range predicted by the
time-bandwidth product, and with significant reduction dependent on
the initial state distribution amongst the subspace defined by the
Hamiltonian eigenvectors. \figref{randH} depicts a typical result with
a random Hamiltonian of dimension $n=2000$. For the three examples
shown, we confined the \black{random} initial states to be \black{linear
  combinations} of 15, 30, and 60 eigenvectors of the Hamiltonian. (Of
course no initial state would be exactly so confined; this illustrates
the effect.) As expected the confinement of the initial state to a
further compressed subspace is seen in the middle plot where the
sorted magnitudes of the unit vector \refeq{eigpsit}
($\aa=V^\dag\psi_0$) clearly drops to zero at exactly 15, 30, and
60. The time-bandwidth product predicts a maximum compression of
$\mtb=32$ which is confirmed by the singular value errors shown in
the lower plot. What is noteworthy is that for the case of the initial
state being confined to a 60 dimensional subspace, the time-bandwidth
product bounds the covariance level compression and the true
covariance error -- the same as $e_{\rm sv}(R)$ -- begins to approach
the sinc matrix error. \black{In other words, the ``sinc'' matrix
  singular values dominate the onset of compression.} A fully random
initial state will cause the covariance singular values to line up
with those of the sinc matrix which bounds the compression. As seen in
the lower plot, as the initial state subspace dimension increases, the
errors get larger, and as expected, do not exceed the sinc matrix
errors.


To emphasize this point, a variety of full length (\ie, $n =2000$)
random initial states were tested, and the compression level matched
that predicted by the time-bandwidth measure.  An interesting point is
that the structured Hamiltonian many-body cases reported earlier in
the paper, and the extreme of random Hamiltonians shown here, both
displayed the same characteristic compression behavior. This situation
indicates that the origin of the compression arises from the
Hamiltonian spectral bandwidth and not any other special features
(\eg, many-body coupling character or patterns).

\section{Further compression with an Autoencoder}
\label{sec:ae}

\begin{figure}[t]
  \centering
 \includegraphics[width=\columnwidth]{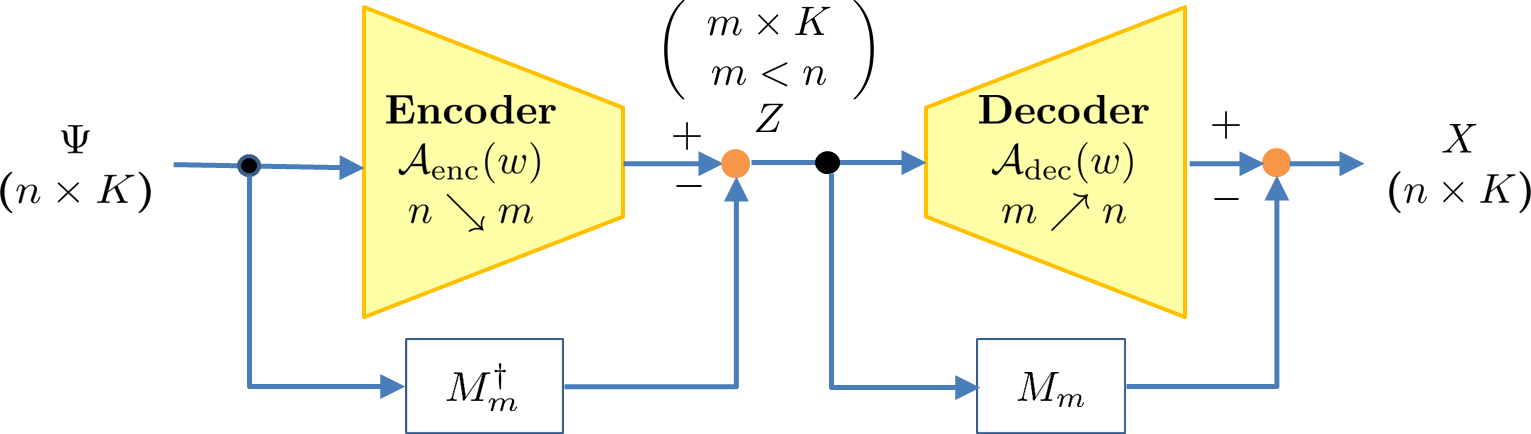}
  \caption{{\bf Autoencoder} with POD pre- and post-processing.}
  \label{fig:autoenc}
\end{figure}

We compare our findings using POD with a variational model constructed
from an autoencoder (AE) as depicted in \figref{autoenc}.  This
configuration is one of several variations for seeking compression by
combining an autoencoder with a data-driven pre-processing procedure
such as POD, \eg, \cite{PanautoPOD:2020}.  By disconnecting the
encoder and decoder the output becomes $X=M_mM_m^\dag\Psi$ which is
exactly the POD solution \refeq{xpod}.

The autoencoder weights $w\in\Rbf^p$ are selected so as to minimize a
weight dependent error, \ie,
\beq[eq:pod ae]
\bea{ll}
\mbox{minimize} &
\ds
E(w) = \normfro{\Psi-X}^2/K,
\\
\mbox{subject to}
&
X = {\cal A}_{\rm dec}(w)(Z)-M_mZ,
\\
&
Z = {\cal A}_{\rm enc}(w)(\Psi)-M_m^\dag\Psi,
\eea
\eeq
where $\normfro{\cdot}$ is the Frobenius norm,
$\Psi=\left[\psi_{t_1}\ \cdots\ \psi_{t_{K}}\right] \in\Cbf^{n\times
  K}$ is the data (snapshot \refeq{snap}) at uniform sample times
$\Del t$. The weight dependent interconnection structure of the
encoder ${\cal A}_{\rm enc}(w):\Cbf^n\to\Cbf^m$, decoder ${\cal
  A}_{\rm dec}(w):\Cbf^m\to\Cbf^n$ and variational parameter dimension
$m<n$ are all specified {\color{black} before employing the AE. That is,
  the AE is seeking to achieve better predictive quality than POD for
  a given $m$ value.  This utilization of the AE (\ie, a special form
  of machine learning algorithm) can be viewed as bringing in a
  nonlinear feature beyond linear use of $M$ as in POD. The flexibility
  inherent in the nonlinear encoding/decoding structure gives the AE
  the potential to produce a smaller state error for the same
  compression level $m$ than that of POD which is restricted to a linear
  encoder/decoder structure.  Finally, unlike in a variational
  simulator, the input state flow is available, embedded here in the
  snap- shot matrix $\Psi$. }

Table \ref{tab:aepod} and \figref{hist podae} compares the POD-only
state errors with those from the POD/AE system. The input data is the
snapshot matrix $\psi\in\Cbf^{1024\times 200}$ from our previous TFI
system example with a random initial state.
The table shows that the POD-only RMS error of 0.5452 at $m=5$ is
reduced to 0.0101 with POD/AE, a value near to that of POD-only for
$m=15$ which has an RMS error as shown of 0.0121. At $m=10,\ 15$ the
RMS errors with POD/AE are approximately those with POD-only at
$m=16,\ 17$, possibly indicating the limit obtainable with this
multi-layer autoencoder. \figref{hist podae} highlights the
large error reduction with the addition of the {\color{black}AE}.

The POD/AE mechanism achieves an error level commensurate with the
time-bandwidth product at $\mtb=14$ for a \comp\ $m=5$ and 10. Such
an error reduction indicates the benefit of the inherent nonlinear
variational AE state. {\color{black}The fundamental reason that the AE,
  and the particular configuration used here, can outperform POD
  remains an open issue.  That is, given that an AE, or any neural
  network, which can approximate most nonlinear functions, what is the
  character of the AE transformation that has this property \ie,
  compression in this case?}

\begin{table}[t]
  \btab{c||c|c}
  $m$ & POD only & POD with AE 
  \\\hline\hline
  5 & 0.5452 & 0.0101 
  \\\hline
  10 & 0.1717 & 0.0070
  \\\hline
  15 & 0.0121 & 0.0030
  \\\hline
  16 & 0.0044 & --
  \\\hline
  17 & 0.0016 & --
  \\\hline
  \etab
  \caption{{\bf Comparison of RMS error} $\normfro{\Psi-X_m}/\sqrt{K}$ from
    POD-only and POD/AE (\figref{autoenc}) for $m=5,10,15$.} 
  \label{tab:aepod}
\end{table}

\begingroup
\renewcommand{\arraystretch}{0.25}
\begin{figure}[t]
  \centering
  \btab{c}  
  \includegraphics[width=\columnwidth]{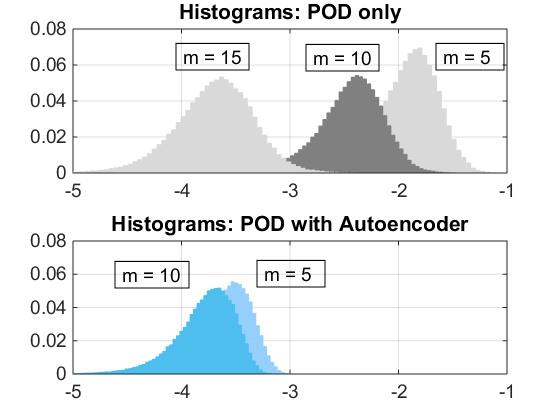}
  \\
  $\log_{10}|(\psi-X)_{ij}|$
  \etab
  \caption{{\bf Histograms} of $\log_{10}$ of the absolute value of all
    elements of the state error matrix $\Psi-X$ \refeq{pod ae}. {\bf
      Upper} POD-only for $m=5,\ 10, 15$. {\bf Lower} POD with
    autoencoder for $m =5,\ 10$.}
  \label{fig:hist podae}
\end{figure}
\endgroup

\section{Compression of Time-Varying Hamiltonian Dynamics}
\label{sec:tv dyn}

\begin{figure}[t]
  \centering
  \includegraphics[width=\columnwidth]{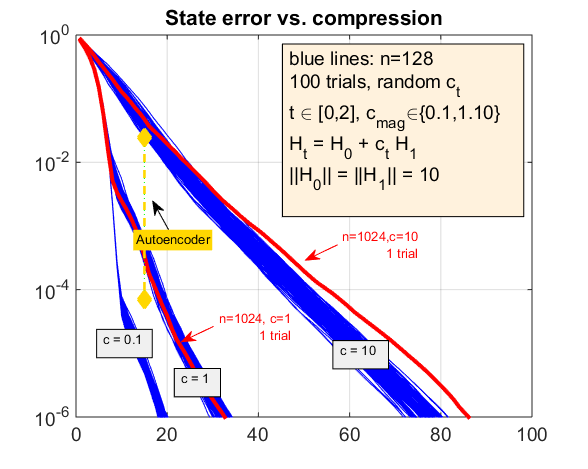}
  \caption{{\bf Linear-Time-Varying random field.} $c_t,t\in[0,2]$. Blue
    curves are POD errors \refeq{sumsvC} for $n=128$ from 100 trials
    each at uniform random control with magnitude ranges $\{\pm
    0.1,\pm 1,\pm 10\}$ from fixed random Hamiltonians normalized as
    indicated. The two red curves are for $n=1024$ from one trial each
    with field magnitude range $\pm 10$ and random Hamiltonians as
    previously normalized. Applying the autoencoder (\figref{autoenc})
    with POD at compression level $m=15, \eps_{15}=0.0245$ (upper
    diamond) results in an error approximately $7\times 10^{-5}$
    (lower diamond).}
  \label{fig:ltv ct}
\end{figure}

Consider the time-varying quantum system,
\beq[eq:ltv ct]
\dot\psi_t = (H_0+c_tH_1)\psi_t,\
\psi_0\in\Cbf^n,\ t\in[0,T],
\eeq
where the external field {\color{black}$c_t$} is independently,
identically, and uniformly distributed in $[-\cmag,\cmag]$ at each of
{\color{black}the} $K$ time intervals in $[0,T]$.  The resulting state
samples are stored in the ``snapshot'' matrix,
\beq[eq:snap tv]
\Psi = \mat{\psi_{t_1}& \cdots & \psi_{t_K}} \in\Cbf^{n\times K},
\eeq
which is used to compute the sampled-data version of the covariance
matrix \refeq{cov} via $C=\Psi\Psi^\dag/K$. The blue curves in
\figref{ltv ct} are the singular value errors $\eps_m(C)$
vs. compression level $m$ from \refeq{sumsvC} for $n=128$ with 100
trials at $K=200$ time samples for $T=2$ and with each field magnitude
$\cmag=\{0.1,1,10\}$ with both $H_0,H_1\in\Cbf^{128\times 128}$
randomly generated, normalized to $\norm{H_0}=\norm{H_1}=10$, and held
fixed throughout the 100 trials. The two red curves are the singular
value errors for one trial with $n=1024$ at the two field magnitudes
$\cmag=\{1,10\}$ and again randomly generated
$H_0,H_1\in\Cbf^{1024\times 1024}$ and normalized also to 10.  We ran
many {\color{black}cases} at $n=1024$; so as not to crowd the figure we
show two representative examples, the rest fell in the same range as the
blue curves. What is interesting to note is that there is compression
in these cases and the levels do not depend very much on the Hilbert
space dimension. They do, however, depend significantly on the
external field magnitude and certainly the relative magnitudes of the
Hamiltonians. Fixing these at 10 for both $n=128$ and $n=1024$ shows
this effect. The results depicted are qualitatively similar to what we
expect for the time-invariant Hamiltonians and the compression level
predicted by the time-bandwidth product.

Though compression is clearly revealed by the POD procedure, it is
based on a \emph{linear} variational state. A nonlinear variational
state, such as one obtained from a neural network would have the
potential for improvement. Applying the autoencoder of
\figref{autoenc} with POD initiated at the compression level $m=15,
\eps_{15}=0.0245$ (upper diamond) results in an error approximately
$7\times 10^{-5}$ (lower diamond), almost a two-order of magnitude
error reduction for the same compression level.

{\color{black} Although a detailed mathematical analysis of the
  time-dependent compression behavior remains to be determined, we
  speculate here} on how a time-bandwidth measure can be developed for
a time-varying quantum system such as \refeq{ltv ct}. Following the
exposition and notation in \cite{BoydLMI:1994}, consider the
\emph{norm-bounded linear differential inclusion } (NLDI),
\beq[eq:nldi]
\bea{rcl}
i\dot\psi_t^{\rm N} &\in& \Hnldi~\psi_t^{\rm N},\ \psi_0^{\rm N}=\psi_0,
\\
\Hnldi &=& \{H_\del=H_0 + \del H_1\ |\ |\del| \leq\cmag \},
\eea
\eeq
where the set $\Hnldi\subset\Cbf^{n\times n}$.
Any solution of \refeq{ltv ct} is also a solution of the NLDI and in
most cases the converse also holds. As a result, many characteristics
of all solutions of the NLDI are inherited by all solutions of
\refeq{ltv ct}.
Our \emph{speculation} is that if compression is one of these
characteristics, then the time-bandwidth product can be applied to the
NLDI to establish the onset level of compression for \refeq{ltv ct}.
Table \ref{tab:nldi} shows the results of using the worst-case
frequency spread from $\Hnldi$ to predict the onset of compression for
\refeq{ltv ct}. Though the predicted NLDI onset of compression is in
the neighborhood produced by POD, absent a more {\color{black}rigorous
  theoretical analysis} we leave this {\color{black}objective} for a
future study.  One possible path we will explore is using approaches
based on robust control theory for multiple uncertainties as applied
in \cite{ssv:2016} for finding the set of uncertain eigenvalues
{\color{black} as might be characterized by the time varying nature of
  a Hamiltonian with multiple time-varying fields.}

\begin{table}[t]
  \btab{|c||c|c|c|}\hline
  \multicolumn{4}{|c|}{\bf Compression Level}\\
    \hline
    $\cmag$
    & \btab{c} NLDI Prediction\\$\max\Del{\rm eig}\{\Hnldi\}$\etab
    & \btab{c}POD error\\$10^{-4}$\etab
    & \btab{c}POD error \\$10^{-5}$\etab
    \\
  \hline
  0.1 & 7 & 10 & 13-15
  \\
  \hline
  1 & 10 & 17-19 & 22-26
  \\
  \hline
  10 & 65 & 44-52 & 58-67
  \\
  \hline
  \etab
  \caption{Comparison of POD compression error levels at $10^{-4}$ and
    $10^{-5}$ with predicted compression onset from the worst-case
    spread of the eigenvalues of $\Hnldi$ \refeq{nldi} for system
    \refeq{ltv ct} with respect to $\cmag\in\{0.1,1,10\}$ for 100 trials
    each for systems evolving from \refeq{ltv ct} with dimension $n=128$
    and $\norm{H_{0,1}}=10$}
  \label{tab:nldi}
\end{table}

\section{Extensions of the analysis}
\label{sec:extend}

In this section we briefly discuss potential extensions of the
time-bandwidth product theory to (A) unitary dynamics and (B) nonlinear
frequency sweeps {\color{black}for constant Hamiltonians}

\subsection{Unitary dynamics}
\label{sec:ucomp}

The time-bandwidth product \refeq{m} also predicts the approximate
size of the compression of a variational simulation of unitary
dynamics,
\beq[eq:udyn]
i\dot U_t = HU_t,\ U_0=I_n,\ t\in[0,T].
\eeq
For $\nu=1,\ldots,n$ let $u_{t,\nu}\in\Cbf^n$ denote the columns of $U_t$.  
Each column of the unitary evolves according to the \emph{same}
Hamiltonian system, \ie,
\beq[eq:ucoldyn]
i\dot u_{t,\nu} = Hu_{t,\nu},\
u_{0,\nu}=\eps_\nu,\ t\in[0,T].
\eeq
Since the initial unitary is the identity matrix, it follows that the initial
value of the $\nu$-th column is $\eps_\nu\in\Rbf^n$, a vector with
single non-zero element equal to one in the $\nu$-th place. Let $X_t$
denote the variational unitary approximation with columns
$x_{t,\nu}\in\Cbf^n,\ \nu=1,\ldots,n$.  Using the Frobenius norm the
state (unitary) error is,
\beq[eq:uerr]
E =\frac{1}{T}\int_0^T\normfro{X_t-U_t}^2~dt
=
\sum_{\nu=1}^n\frac{1}{T}\int_0^T\norm{x_{t,\nu}-u_{t,\nu}}^2_2~dt.
\eeq
As shown in Appendix \ref{sec:unitary}, under a linear sweep of
Hamiltonian eigenvalues, the previously defined sinc matrix bounds all
the singular values of each subsystem covariance resulting in the
total error bound,
\beq[eq:Elin]
E \approx E_\lin \leq n\left(1-\sum_{i=1}^m\sv_i(\Slin/n)\right).
\eeq
As we have shown for compression of the quantum state dynamics with a
constant Hamiltonian, we see the same here, \ie, for large $n$ the
singular values of $S_\lin$ approach those of $\Shlin$, and as in
\refeq{esvSlin}, the error tends asymptotically to zero, thereby
ensuring compression for a value of $m$ not dependent on the
exponential growth of the Hilbert space dimension with the number of
particles.

\subsection{\color{black}Nonlinear frequency sweeps}
\label{sec:Beyond}

\begin{figure}[t]
  \centering
  \btab{cc}
  \includegraphics[width=0.5\columnwidth]{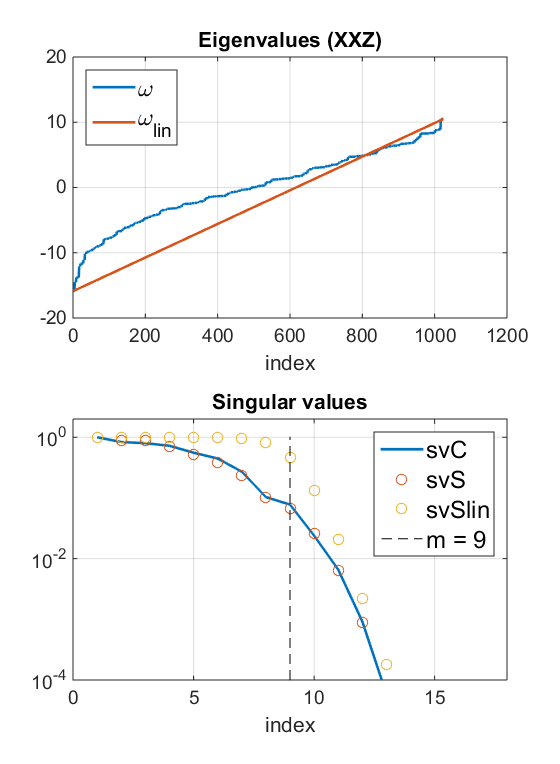}
  &
  \includegraphics[width=0.5\columnwidth]{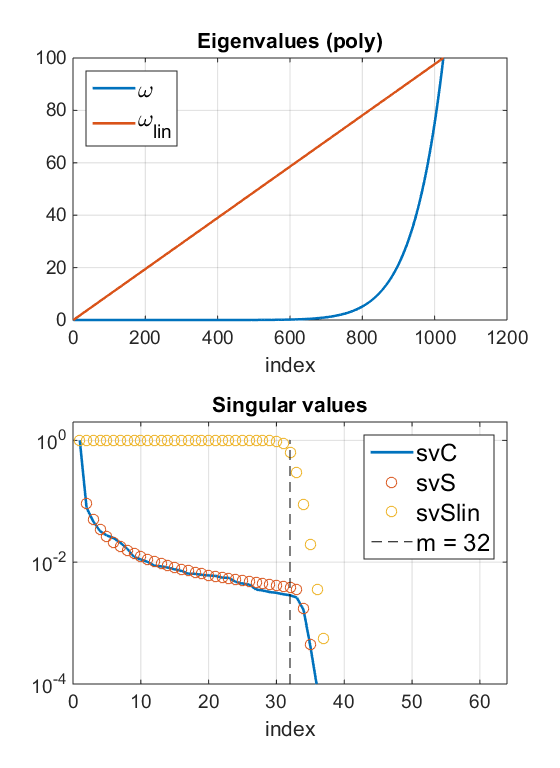}
  \\
  \bf{(a)} & \bf{(b)}
  \etab
  \caption{{\bf Two types of eigenvalue distributions} with system
    dimension $n=1024$. {\bf (a)} from an XXZ spin-chain exhibiting
    characteristic piecewise constant variations, and {\bf (b)}
    fictitious system with an exaggerated polynomial distribution.}
  \label{fig:eigs}
\end{figure}

The more common case to expect from an arbitrary Hamiltonian is a
nonlinear eigenvalue sweep (\eg, see Figure~\ref{fig:htfi}).
\figref{eigs} presents two other types of eigenvalue spreads and the
corresponding singular value plots of the covariance matrix $C$, the
sinc matrix $S$, and the linear sinc matrix $\Slin$. The upper plot in
{\bf(a)} shows the eigenvalues from an XXZ spin-chain with Hamiltonian
$H=\sum_i
\sig_x^i\sig_x^{i+1}+\sig_y^i\sig_y^{i+1}+\del\sig_z^i\sig_z^{i+1}$. The
characteristic feature of these eigenvalues is that they are piecewise
constant over varying intervals. With $\del=1.5$ and with a random
initial state, the lower plot in {\bf(a)} shows that the linear
approximation gives a compression that is a comparable range with that
obtained from POD of a snapshot matrix. On the right the upper plot in
{\bf(b)} shows a fictitious eigenvalue distribution made from an
exaggerated polynomial where the eigenvectors are selected randomly to
generate the Hamiltonian. Though the linear approximation is clearly
not very good, the compression estimate {\color{black}remains} in the
range predicted by the time-bandwidth product.

Though these circumstances violate the assumption leading to the
time-bandwidth product analysis, the examples presented earlier in the
paper show that a nonlinear eigenvalue sweep does not result in a
serious problem: the predicted compression from the linear eigenvalue
sweep is in the range computed from the actual system. To further
support this finding we propose to “stretch” the nonlinear eigenvalue
sweep engendered by an $n$-dimensional Hamiltonian into a
“straightened out” linear sweep over the same range. This is
accomplished by equating the path lengths of the two sweeps.

The path length of $n$ nonlinear monotonically increasing eigenvalues
from $\om_1$ to $\om_n$ is,
\beq[eq:len]
d(\om) = \sum_{k=1}^{n-1}
\left((\om_{k+1}-\om_k)^2 + 1 \right)^{1/2}.
\eeq
If the sweep were linear then $d(\om_{\rm
  lin})=\sqrt{(\Del\om)^2+(n-1)^2}$ with $\Del\om=\om_n-\om_1$. To
keep the same eigenvalue range with the stretched linear sweep
requires interpolating between the stretched gaps in successive
eigenvalues with an additional $n'-n$ eigenvalues where,
\beq[eq:nlen]
n' = \lceil 1 + \left( d(\om)^2 - (\Del\om)^2\right)^{1/2} \rceil
\eeq
For the two TFI frequency sweeps shown in \figref{htfi} for
$n\in\{1024,4096\}$ the stretched system dimension increases modestly
to $n'\in\{1029,4101\}$. Since the \comp\ $m$ scales logarithmically
with dimension \refeq{mlogn}, the relative time-bandwidth \comp\
estimate is $\log n'/\log n-1\in\{1.47\times 10^{-4},7.03\times
10^{-4}\}$. These changes are not visible in \figref{svSSlin} where
the predicted \comps\ under the linear sweep assumption are compared
with the order where the actual singular values become
insignificant. Similarly for the $XXZ$ and polynomial eigenvalues
shown in \figref{eigs}: $n'_{\rm xxz}=1046$ and $n'_{\rm
  poly}=1047$.



\section{Summary}\label{sec:sum}

The main result reported here is the introduction of the
time-bandwidth product,
\beq[eq:tbw]
\delf=(f_{\max}-f_{\min})T
\eeq
which, for a quantum state evolving for time $T$ with a constant
Hamiltonian whose frequencies range from $f_{\min}$ to $f_{\max}$,
provides an estimate of compression to a reduced {\color{black}system}
whose dimension is the nearest integer. We show that the predicted
compressed dimension is exact (in a well defined asymptotic sense) if
the Hamiltonian frequencies (eigenvalues) range linearly when ordered
from minimum to maximum.  Since the time-bandwidth product does not
depend on the initial state, it is, in effect, predicting the range of
the worst-case level of compression, or more precisely, where
compression begins as defined in the Introduction. Though every real
system has a nonlinear frequency sweep, numerous simulations of
systems with random (constant) Hamiltonians shown here do not violate
the predicted approximate level of compression, \ie, no orders of
magnitude changes.  In general, the time-bandwidth product is
consistent with the level or onset of observed compression.

The time-bandwidth product is of course not independent of the Hilbert
space dimension. For spin systems composed of two-level particles, the
frequency range typically scales linearly with the number of
particles. As a result the time-bandwidth estimated compression level
scales logarithmically. More generally a polynomial scaling of
frequency range will still result in a compression that beats
exponential scaling.

A lower value of the predicted level of \comp, and sometimes
significantly lower, is possible dependent upon how the initial state
is distributed amongst the Hamiltonian eigenvectors.  {\color{black}In
  particular, for a system with a random initial state, the system
  dimension has little impact on the variational compression, the main
  driver being the product of the frequency sweep range and the
  simulation run time. In contrast, for a specific initial state, such
  as one close to the system ground state, the interaction with the
  Hamiltonian eigenvectors plays a significant role in further
  reducing the variational compression. } We showed how this comes
into play using a related measure: if the initial state is
predominantly in or near a subspace spanned by a small number of
Hamiltonian eigenvectors, compression will be in the range of the
initial subspace dimension, lower than that indicated by the
time-bandwidth product.  With enough run-time, and with the initial
state not predominantly confined to a subspace, all of Hilbert space
will eventually be populated. The good news is that this upper
dimension can only grow linearly with run-time: with a huge number of
states the run-time would have to be very long to make such an
impact~\cite{illusion:11}.

A limitation of the time-bandwidth product is that it is derived from
a \emph{linear} variational model. In contrast, a machine learning
system is built to implement a \emph{nonlinear} transformation, and
thereby potentially delivering a lower level of compression with the
same or smaller error. We {\color{black}demonstrated} this effect using
an autoencoder. On the other hand, the time-bandwidth product does
reveal the \emph{existence} of a useful level of compression; finding
that with a machine learning system {\color{black}to actually solve for
  the dynamics is an evolving area of challenging research}.

\black{For a time-varying Hamiltonian, \eg, a system affected by
  time-varying external fields, the time-bandwidth product is not
  strictly applicable. Nevertheless, simulations show that compression
  still holds, though with an expected dependence on the field
  strength. It also seems reasonable to expect compression with a
  time-varying Hamiltonian to be worse than the time-invariant case,
  as the external field can be thought of as moving the initial state
  around through some portion of the specified Hilbert space. For a
  reasonably posed control problem, even for an exponentially large
  many-body system, it would normally not be expected to define the
  goal for the control to move from one end of Hilbert space to the
  other.}

The time-bandwidth predicted compression in some ways reveals that the
\schrod equation is a giant variational minimization machine within
which is found a ``discovery:'' \emph{Compression}.  Returning to
the second paragraph of the paper with respect to points (1)and (2)
there, one can also view the system compression as an asymptotic
result, where the formulation presented in the paper and the numerical
evidence clearly indicates that a tolerable level of the onset of
compression appears to typically set in at very low values of $m$ with
$m \ll n$.  That possibility is of course buried in the data that goes
into any machine learning system. It would seem, then, that a neural
net quantum simulator would have to find a compressed system, or else
how could it simulate quantum dynamics (with usefully small errors)
without using a number of parameters equivalent to the exponential
size of the quantum state. Though we have no proof at this time, the
reports of successful simulations of quantum many-body dynamics with
non-exponentially scaling of neural network parameters lend support to
the time-bandwidth product prediction of the existence of compression.


\textbf{Acknowledgments} {\color{black}All of the authors acknowledge
  support by the Data X project at Princeton University.}  RLK partly
supported under the Defense Advanced Research Projects Agency (DARPA)
Physics of Artificial Intelligence (PAI) Program (Contract
HR00111890031). RLK thanks Shaowu Pan for alerting us to the POD
modified autoencoder structure and Jun Kyu Lee and Kamal Nayal for the
implementation and data assembly \black{thereof}.



\bibliographystyle{unsrt}
\bibliography{rlk,Qcompress-1} 

\begin{appendix}

\section{Singular values of $R$}
\label{app:svR}

Using the eigenvalue decomposition of the Hamiltonian and state flow
\refeq{eigH}-\refeq{eigpsit}, the covariance matrix \refeq{cov} can be
expressed as,
\beq[eq:covF]
C[\psi] = VFV^\dag,\quad
F = \frac{1}{T}\int_0^T f_t f_t^\dag~dt \in\Cbf^{n\times n}.
\eeq
Because $V$ is unitary, the singular values of the state covariance
matrix $C$ (we drop the $C[\psi]$ notation for clarity) are identical
to those of $F$:
\beq[eq:svdF]
\bea{c}
\left(
\bea{l}
F = W\mat{Q_m&0\\0&Q_{n-m}}W^\dag,
\\
W = \mat{W_m&W_{n-m}},\ W_m\in\Cbf^{n\times m}
\eea
\right),
\\
\Downarrow
\\
\left(
\bea{l}
\mat{M_m&M_{n-m}} = VW
\\
M_m = VW_m\in\Cbf^{n\times m}
\eea
\right).
\eea
\eeq
The matrices $Q_m,Q_{n-m}$ are diagonal and contain the singular
values of $C$ (or $F$) in descending order (same as in
\refeq{cov}). Since $V,W\in\U(n)$, the singular vectors of $C$ are the
columns of the unitary product $VW\in\U(n)$ from which the first $m$
columns provide the basis for the variational model reduction
corresponding to the model error $\trace~Q_{n-m}$ \refeq{Epodmin}.
An equivalent expression for $F$ is,
\beq[eq:FRhat]
\bea{l}
F = \diag(e^{-i\om T/2})~R~\diag(e^{-i\om T/2})^\dag,
\\
R = \diag(\aa)~S~\diag(\aa)^\dag,
\eea
\eeq
with $R$ and $S$ as defined in \refeq{RSdecomp}-\refeq{RSkell}.
Since $\diag(e^{-i\om T/2})$ is a unitary, the singular values of $F$ are
the same as those of $R$ and also $C$. 

\section{Singular values of $\Slin$}
\label{app:svS}


\begin{figure}[t]
\centering
  \btab{c}
  \includegraphics[width=\columnwidth]{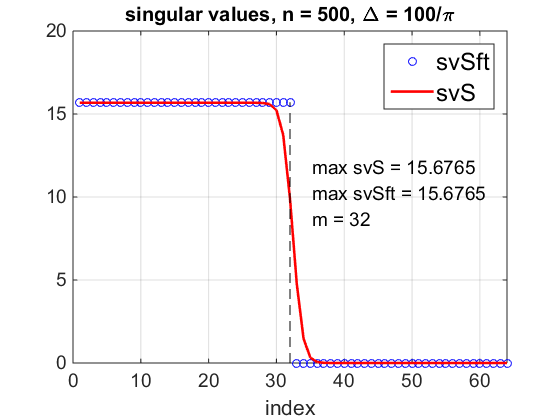}
  \\{\bf(a)}\\
  \includegraphics[width=\columnwidth]{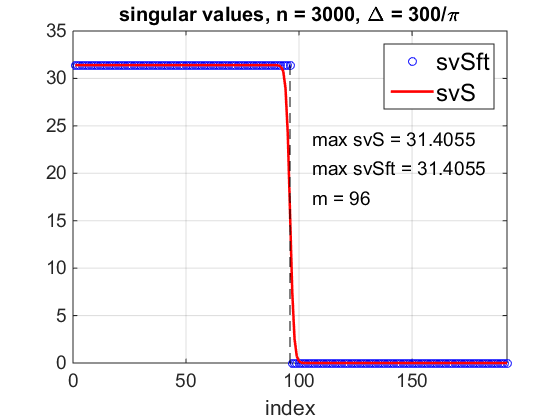}
\\{\bf(b)}
  \etab
  \caption{{\bf Singular values} of ideal \refeq{svSlin} and $\Slin$.
    {\bf(a)} $\{n=500,\delf=100/\pi\}$ and {\bf(b)} $\{n=3000,\delf=300/\pi\}$
    with \comp\ estimate $\mtb=
    \Big\lceil\left(\frac{n}{n-1}\right)\delf\Big\rceil$ from
    \refeq{m}.}
\label{fig:svSSlinA}
\end{figure}

Let $T_L\in\Rbf^{L\times L}$ be a real symmetric Toeplitz matrix whose
first column is generated by the series $\bar
f_L=\{f_\ell,\ell\in[0,L-1]\}$.  From \cite{Bottcher:2017}, if the
Fourier Transform $F(\om)={\cal F}(\bar f_\infty)$ satisfies certain
monotonicity conditions, then as $L\to\infty$ the eigenvalues of $T_L$
approach those of the Fourier transform $F(\om)$. The first column of
the sinc-matrix with a linear sweep $\Slin$ \refeq{delf} are the first
$n$ terms of the of the $L$-length series $\bar
s_L=\{\sinc\left(\frac{\pi\delf\ell}{n-1}\right),
\ell\in[0,L-1]\}$. With $n$ fixed, the Fourier Transform of $\bar
s_\infty$ satisfies the aforementioned conditions, \ie, from standard
tables,
\beq[eq:eigSlin]
\bea{rcl}
F(\om) &=& \sum_{\ell=-\infty}^\infty e^{-i\om\ell}
~\sinc\left(\frac{\delf\pi\ell}{n-1}\right)
\\
&=& \left\{
\bea{ll} (n-1)/\delf, & |\om| \leq \pi\delf/(n-1),
\\
0, & \pi\delf/(n-1) < |\om| < \pi.
\eea
\right.
\eea
\eeq
Discretizing $\om=2\pi k/n,\ k=0,\ldots,n-1$, and then sorting and
grouping the absolute values of $|F(\om_k)|$, we get the singular
value (asymptotic in $n$) approximation \refeq{svSlin}. (In
\cite{Ekstrom:17} this type of asymptotic approximation was used for
Toeplitz eigenvalues.) \figref{svSSlinA} compares the ideal
(Fourier transform) singular values \refeq{svSlin} labeled {\bf svSft}
with those from the linear sweep matrix $\Slin\in\Rbf^{n\times n}$
labeled {\bf svS} for two instances: (a) $\{n=500,\delf=100/\pi\}$ and
(b) $\{n=3000,\delf=300/\pi\}$ with \comp\ estimate $\mtb=
\Big\lceil\left(\frac{n}{n-1}\right)\delf\Big\rceil$ from \refeq{m}.

\section{Unitary Compression}
\label{sec:unitary}

Applying the POD method, \emph{mutatis mutandis}, to each of the $n$
systems \refeq{ucoldyn} with the variational \comp\ fixed for all
at $m$, results in the optimal \emph{linear} variational unitary as,
\beq[eq:var unitary]
X_t = \mat{\Gam_1u_{t,1} & \cdots & \Gam_nu_{t,n}},
\eeq
where the rank-$m$ matrices $\Gam_\nu=M_\nu M_\nu^\dag\in\Cbf^{n\times
  n}$ with each $M_\nu\in\Cbf^{n\times m}$ formed as in \refeq{cov}
from the $m$ singular vectors corresponding to the $m$ largest
singular values of each covariance matrix,
\beq[eq:cov nu]
C_\nu = \frac{1}{T}\int_0^T u_{t,\nu}u_{t,\nu}^\dag~dt.
\eeq
The optimal (POD) state (unitary) error and the corresponding $R$ and
$S$ matrices are,
\beq[eq:unitary error]
\bea{rcl}
E &=&
\ds
\sum_{\nu=1}^n\trace(I_n-\Gam_\nu)C_\nu,
\\
C_\nu &=& V\diag(e^{-i\om T/2})~R_\nu~\diag(e^{i\om T/2})V^\dag,
\\
R_\nu &=& \diag(\alf_\nu)~S~\diag(\alf_\nu)^\dag,\
\alf_\nu=V^\dag\eps_\nu.
\eea
\eeq
Here $S$ is \emph{exactly} the previously defined sinc matrix
\refeq{RSdecomp}. The error bound \refeq{Elin} follows directly.

\clearpage


\end{appendix}

\end{document}